\documentclass[prd,aps,showpacs,preprintnumbers,amsmath,amssymb,11pt]{revtex4}
\usepackage{graphicx}
\usepackage{dcolumn}
\usepackage{bm}
\usepackage{psfrag}

\begin{document}
\newcommand{\be}{\begin{equation}}\newcommand{\ee}{\end{equation}}
\newcommand{\bea}{\begin{eqnarray}}\newcommand{\eea}{\end{eqnarray}}
\newcommand{\bc}{\begin{center}}\newcommand{\ec}{\end{center}}
\def\no{\nonumber}
\def\eq#1{Eq. (\ref{#1})}\def\eqeq#1#2{Eqs. (\ref{#1}) and  (\ref{#2})}
\def\lsim{\raise0.3ex\hbox{$\;<$\kern-0.75em\raise-1.1ex\hbox{$\sim\;$}}}
\def\gsim{\raise0.3ex\hbox{$\;>$\kern-0.75em\raise-1.1ex\hbox{$\sim\;$}}}
\def\slash#1{\ooalign{\hfil/\hfil\crcr$#1$}}
\def\eff{\mbox{\tiny{eff}}}
\def\order#1{{\mathcal{O}}(#1)}
\def\pppm{B^0\to\pi^+\pi^-}
\def\pzpz{B^0\to\pi^0\pi^0}
\def\pppz{B^0\to\pi^+\pi^0}
\preprint{052010}
\title{$\eta-\eta^{\prime}$ Mixing Angle from Vector Meson Radiative Decays}
\author{
T. N. Pham}
\affiliation{
 Centre de Physique Th\'{e}orique, CNRS \\ 
Ecole Polytechnique, 91128 Palaiseau, Cedex, France }
\date{\today}
\begin{abstract}
The octet-singlet  $\eta -\eta^{\prime}$ mixing mass term could have a
derivative  $O(p^{2})$ term  as found in recent analysis of the 
$\eta -\eta^{\prime}$ system. This term gives rise to an additional 
momentum-dependent pole contribution which is 
suppressed by a factor  $m_{\eta}^{2}/m_{\eta^{\prime}}^{2}$ for $\eta$
relative to the  $\eta^{\prime}$ amplitude. The processes with 
$\eta$ meson can then  be described, to a good approximation, 
by the momentum-independent mixing mass term which gives rise to 
a new $\eta -\eta^{\prime}$ mixing angle $\theta_{P}$, 
like the old $\eta -\eta^{\prime}$ 
mixing angle used in the past, but a  momentum-dependent mixing term 
$d$, like   $\sin(\theta_{0}-\theta_{8})$ in 
the two-angle mixing scheme used in the parametrization of the 
pseudo-scalar meson decay constants in the current literature,  is needed 
to describe the amplitudes with $\eta^{\prime}$. In this paper, we obtain
sum rules relating  $\theta_{P}$ and  $d$ to the physical vector meson 
radiative decays with $\eta$ and $\eta^{\prime}$, as done in our previous 
work for  $\eta$ meson two-photon decay, and with nonet symmetry for
the $\eta^{\prime}$ amplitude, we obtain a mixing angle 
  $\theta_{P}=-(18.76\pm 3.4)^{\circ}$,  $d=0.10\pm 0.03$ from
$\rho\to\eta\gamma$  and $\eta^{\prime}\to\rho\gamma$ decays, for 
$\omega$ ,  $\theta_{P}=-(15.81\pm 3.1)^{\circ}$, 
$d=0.02\pm 0.03$, and for $\phi$,  $\theta_{P}=-(13.83\pm 2.1)^{\circ}$, 
$d=0.08\pm 0.03$. A larger value of $0.06\pm 0.02$ for $d$ is obtained 
directly from the nonet symmetry 
expression for the $\eta^{\prime}\to\omega\gamma$ amplitude. This 
indicates that more precise  vector meson radiative decay measured
 branching ratios
and higher order $SU(3)$ breaking effects could bring these  
values for $\theta_{P}$ closer and allows a better determination of $d$.

\end{abstract}
\maketitle

The $\eta -\eta^{\prime}$ mixing angle plays an important role in physical
processes involving the light pseudo-scalar meson nonet, the $\eta$
and $\eta^{\prime}$ mesons. In the presence of $SU(3)$ breaking due to
the large current s-quark mass compared to the light u and d current quark 
mass, with $m_{s}\gg m_{u,d}$, the octet $\eta_{8}$ and the singlet
$\eta_{0}$ could mix with each other through a small $SU(3)$ symmetry
breaking quark mass term and generate the two physical states, the $\eta$
and $\eta^{\prime}$. Since $m_{s}\ll\Lambda_{\rm QCD}$ , and 
because of the $U(1)$ QCD-anomaly, the $\eta_{0}$
mass is much larger compared to the $\eta_{8}$ mass, the 
$\eta-\eta^{\prime}$ mixing angle  is 
$O(m_{s}/\Lambda_{\rm QCD})$ so that
the physical $\eta$ and $\eta^{\prime}$ are almost  pure
$\eta_{8}$ and $\eta_{0}$ eigenstate respectively, in contrast with the ideal
mixing for the $1^{-}$ low-lying vector meson states. Assuming nonet 
symmetry for the off-diagonal mass term $<\eta_{0}|H_{\rm SB}|\eta_{8}>$,
one would get a mixing angle $\theta_{P}= -18^{\circ}$ \cite{Kawarabayashi}
in good agreement with
the  value $\theta_{P} \approx -(22 \pm 3)^{\circ}$ in \cite{Donoghue}, or 
$\theta_{P} \approx -(18.4 \pm 2)^{\circ}$ in \cite{Pham} obtained from 
the  $\eta$ and $\eta^{\prime}$ two-photon width. The
large mixing angle obtained from the two-photon decay rate 
is consistent with  nonet symmetry \cite{Pham1} (mixing angle with
linear Gell-Mann-Okubo(GMO) mass formula is given in \cite{Gilman,Morpurgo}).  A previous
phenomenological  analysis many years ago \cite{Gilman} already found     
a large mixing angle $\theta_{P}\approx -(20-23)^{\circ}$ in 
the  pseudo-scalar meson two-photon widths, in  $J/\psi \to
\gamma\eta(\eta^{\prime})$,  $J/\psi \to V P$, in radiative decays 
of light vector mesons, and in   $\pi^{-}p$ scattering a mixing angle 
$\approx -20^{\circ}$ is favored, but  
light tensor meson decays seem to favor a mixing angle of 
$\approx -10^{\circ}$, the GMO mass formula value.
 Subsequently, a value between $-13^{\circ}$ and $-17^{\circ}$, or an
average $\theta_{P}= -15.3^{\circ} \pm 1.3^{\circ}$ is obtained 
 \cite{Bramon} and  
$\theta_{P}\approx -11^{\circ}$  is obtained in \cite{Benayoun}. 
Recent analysis \cite{Escribano,KLOE} using the more  precise
$V\to P\gamma$ measured branching ratios \cite{PDG} found a mixing angle
$\theta_{P}=-13.3^{\circ} \pm 1.3^{\circ}$. It appears that the mixing 
angle obtained in these recent theoretical calculations 
is a bit smaller  than 
the nonet symmetry value\cite{Kawarabayashi} (the nonet
symmetry value is very close to the  mixing angle value we obtained from
$\eta$ two-photon decay rate using only the measured $\eta^{\prime}$
two-photon decay rate\cite{Pham}). This could be due to various
 theoretical uncertainties,  like the use of nonet
symmetry in the treatment of  radiative decays involving 
 $\eta^{\prime}$  and possibly, the neglect of  higher order $SU(3)$ 
breaking in the radiative decay amplitudes. Also, most of the analysis in
the past is based on the assumption that the off-diagonal octet-singlet 
transition mass term does not depend significantly on the energy of the state
\cite{Donoghue1}. Recent works\cite{Leutwyler,Kaiser,Feldmann} 
show that a quadratic derivative off-diagonal octet-singlet transition of the
form $\partial_{\mu}\eta_{0}\,\partial_{\mu}\eta_{8}$
requires two angles $\theta_{8}$ and $\theta_{0}$ to describe the
pseudo-scalar meson decay constants. Here we adopt a simple approach to
describe the $\eta-\eta^{\prime}$ system. We consider the 
$\eta-\eta^{\prime}$ system with 
the  non-derivative off-diagonal mass term  diagonalized by the 
usual mixing angle $\theta_{P}$ and  the additional 
off-diagonal derivative  $SU(3)$ breaking mass term   treated as a 
perturbation:
\be
{\cal L}_{\rm SB}  = d\,\partial_{\mu}\eta_{0}\,\partial_{\mu}\eta_{8}
\label{L8}
\ee
where $d$ is first order in $SU(3)$ breaking parameter(
$O(m_{s}/\Lambda_{\rm QCD})$).
The two $\eta$ and $\eta^{\prime}$ physical states are still the usual
linear combinations of the pure  singlet and octet $SU(3)$ state with the
monentum-independent mixing angle $\theta_{P}$, but the momentum-dependent 
off-diagonal mass term will give rise to an additional contribution to 
processes involving $\eta$ and $\eta^{\prime}$ by the quadratic momentum
dependent pole term( as  in non-leptonic $K\to 3\pi$ decays \cite{Cronin}, 
for which the $K$ meson pole term is suppressed relative  to the pion
pole term by the factor $m_{\pi}^{2}/m_{K}^{2}$). 
The $\eta^{\prime}$ pole contribution to the process with $\eta$ on the 
mass shell is of the strength $d\,(m^{2}_{\eta}/m^{2}_{\eta^{\prime}})$,
a second order $SU(3)$ breaking effect and is 
suppressed by the factor $ m^{2}_{\eta}/m^{2}_{\eta^{\prime}}$. 
The $\eta$ pole contribution to the  $\eta^{\prime}$ amplitude is of
a strength $d$,  a  first order $SU(3)$ breaking mixing term, like 
the $\sin\theta_{P}$ term.  Thus the 
quadratic momentum-dependent off-diagonal  mixing mass term, while 
leaves the amplitude with $\eta$ almost unaffected, could enhance or 
suppress the $\eta^{\prime}$ amplitude. This seems to be the 
origin of the two-angle description
of the pseudo-scalar meson decay constants introduced in the literature as
mentioned above. The angle  $\theta_{8}$, like the new mixing angle
in our scheme ( denoted by $\theta_{P}$ in the following), behaves 
like the old mixing angle and effectively describes the mixing 
of  $\eta_{0}$ with $\eta_{8}$ to make the physical $\eta$ meson  
while $\sin\theta_{0}$ would 
effectively give the admixture of $\eta_{8}$
in $\eta^{\prime}$. 
There have been recent calculations of 
vector  meson radiative decays\cite{Benayoun1,Escribano1}, using 
the two-angle mixing scheme with the result that the angle $\theta_{8}$
is  quite  close to the nonet symmetry value in the one-mixing angle 
analysis, while $\theta_{0}$ is found to be rather small, implying a 
smaller admixture of the $\eta_{8}$
component in $\eta^{\prime}$ than  the case with one mixing angle . If 
one neglects second order in $SU(3)$
breaking parameters, the determination of the old and new mixing angle
would give essentially the same result and the results for the mixing
angle obtained in the past still apply, in particular our previous
result from the two-photon $\eta$ meson  decay rates\cite{Pham}. We now
apply our method to  vector meson radiative decays to obtain 
first $\theta_{P}$ with the  sum rules for $\eta$  and then determine
both $\theta_{P}$ and $d$ using both sum rules for $\eta$ and $\eta^{\prime}$
and  nonet symmetry for the pure singlet 
$V\to \eta_{0}\gamma$ amplitude. The  sum rules for $\eta$
gives  a mixing angle in the range $-(14-17)^{\circ}$, while the two
sum rules give  similar value $\theta_{P}$ in the range
$-(14-19)^{\circ}$ 
and a  value for $d$  in the range $0.08-0.10$ for 
$\rho,\phi$ radiative decays, but a very small  $d=0.02$ for $\omega$ decay. 

 Since the $\eta-\eta^{\prime}$ mixing is an additional
$SU(3)$ breaking effect not present in the decay amplitude for the pure octet
$\eta_{8}$  state, the difference between the decay
involving the physical $\eta$ meson and the  $\eta_{8}$ state is a measure
of  the $SU(3) $ octet-singlet mixing effect, it is thus possible to express 
this difference in terms of the measured radiative decay branching ratios and
a minimum theoretical input without involving the pure singlet $\eta_{0}$
state. This method has been used in a 
determination of  the $\eta-\eta^{\prime}$ mixing angle 
without involving  the pure singlet $\eta_{0}\to \gamma\gamma$ amplitude 
which is usually obtained with nonet symmetry. We have, without the
momentum-dependent mixing mass term:
\be
A_{\eta}\cos \theta_{P} + A_{\eta^{\prime}}\sin \theta_{P}  = \frac{f_{\pi}} {f_{\eta_{8}}}(1 - \delta)\frac{A_{\pi}} {\sqrt{3}}
\label{etagg}
\ee
where $\delta= -0.27$ as estimated in \cite{Pham} from the  continuum 
contribution of the $SU(3)$ breaking effects to the anomaly term,
similar to $SU(2)$ breaking terms for two-photon $\pi^{0}$ 
decay \cite{Kitazawa}. The expressions with the momentum-dependent 
$\eta-\eta^{\prime}$ transition  included are obtained by making a 
substitution in Eq. (\ref{etagg}) :
\bea
&&A_{\eta} \to A_{\eta}
+d\,(m^{2}_{\eta}/m^{2}_{\eta^{\prime}})A_{\eta^{\prime}}, \nonumber \\
&&A_{\eta^{\prime}} \to A_{\eta^{\prime}}-d\,A_{\eta}. 
\eea
These additional mixing terms will contribute to the 
l.h.s of  Eq. (\ref{etagg}) terms second order
in $SU(3)$ breaking parameters. Since second order in $SU(3)$
breaking in the r.h.s of Eq. (\ref{etagg}) is not known at present, 
for example, in the two-angle mixing scheme, the  quantity 
$\sin(\theta_{0}-\theta_{8})$ is given to leading order in $SU(3)$ 
breaking mass term \cite{Kaiser}, to be
consistent, one has to drop all second order terms in the Eq. (\ref{etagg}). 
This allow a determination of the new mixing angle from the measured
pseudo-scalar two-photon and vector meson radiative decays without large  
theoretical uncertainties which could be due to possible second order $SU(3)$
breaking terms in vector meson radiative decays. This seems to be the price
to pay for the presence of the momentum-dependent mixing mass term
which now should be determined from the amplitude with $\eta^{\prime}$.
This is also the reason to use the sum rules in Eq. (\ref{etagg}) 
which involves  only the measured decay rates with 
$\eta$ and $\eta^{\prime}$. 

The above sum rules shows clearly that the difference between the
physical $\eta$ and the pure $\eta_{8}$ two-photon decay amplitude 
is a direct measure of the mixing effect
and hence give us the mixing angle using only the
measured $\eta^{\prime}$ two-photon decay rate. Since our purpose is to
extract only the mixing angle and not to make a theoretical calculation
of  $\eta^{\prime}\to \gamma\gamma$, we do not need a theoretical  
expression for the pure $\eta_{0}$ two-photon decay amplitude.  
Eq. (\ref{etagg}) gives \cite{Pham}
\be
\theta_{P} = -(18.4 \pm 2)^{\circ} 
\label{theta}
\ee
which is also practically the value obtained with  the current
measured $\eta\to \gamma\gamma$ branching
ratio \cite{PDG} which has not changed over the years ($\theta_{P} =
-(18.1 \pm 2)^{\circ} $) with the current data. This value is 
in good agreement with the nonet symmetry value of $-18^{\circ} $
obtained with the first order SU(3) breaking mass term 
in \cite{Kawarabayashi}. This shows that at least to first order in $SU(3)$
breaking, one can use Eq. (\ref{etagg}) to determine the new mixing angle.
We now apply this method to extract the $\eta-\eta^{\prime}$ 
mixing angle from radiative
decays of light vector mesons $V\to P\gamma$.
In addition to  $SU(3)$  and nonet symmetry breaking effects
in the magnetic coupling for  $V\to\eta_{8}\gamma$ and
$V\to\eta_{0}\gamma$ amplitude, there
is also an $SU(3)$ and nonet symmetry breaking $O(p^{2})$ derivative
coupling term which requires a renormalization of $K$ meson, $\eta_{8}$
and $\eta_{0}$ field operator \cite{Pham2,Gasser} 
by the factor $f_{\pi}/f_{K}$, $f_{\pi}/f_{\eta_{8}}$ and 
$f_{\pi}/f_{\eta_{0}}$ to put the propagator in the canonical 
$\frac{1}{(p^{2}-m^{2})}$ form.
Given these   $SU(3)$ and nonet symmetry breaking effects, similar 
expressions like Eq. (\ref{etagg}) for $V\to \eta,\eta^{\prime}\gamma$, 
$V=\rho,\omega,\phi$ are
obtained and the $\eta-\eta^{\prime}$ mixing angle can be determined 
in a very simple manner. 

Let $|\eta_{0}>, |\eta_{8}>$ be the two $SU(3)$ singlet and octet
states   of the  pseudo-scalar $I=0$  $SU(3)$ nonet in terms of the
flavor diagonal $q \bar{q}$ component:
\bea
&& |\eta_{0} > =  (| u \bar{u}+ d \bar{d} + s \bar{s}>)/\sqrt{3} , \nonumber\\
&&  |\eta_{8} > = (| u \bar{u}+ d \bar{d} - 2\,s \bar{s}>)/\sqrt{6} .  
\label{eta08}
\eea
In the presence of  $SU(3)$ symmetry breaking quark mass term, the 
mixing of  $\eta_{0}$ with  $\eta_{8}$ will produce
the two physical states,  $\eta$ and $\eta^{\prime}$ which are  given by the
linear superpositions of the pure $\eta_{0}$ and $\eta_{8}$ states
obtained by an unitarity transformation to diagonalize the mass matrix.
\bea
&& |\eta > = \cos\theta_{P}|\eta_{8} > - \sin\theta_{P}|\eta_{0} > , \nonumber\\
&&  |\eta^{\prime} > = \sin\theta_{P}|\eta_{8} > +\cos\theta_{P}|\eta_{0} > .  
\label{eta}
\eea
in terms of the mixing angle $\theta_{P}$. By inverting Eq. (\ref{eta})
one can express  $\eta_{0}$ and $\eta_{8}$ states
in terms of the physical states $\eta$ and  $\eta^{\prime}$ as:
\bea
&& |\eta_{8} > = \cos\theta_{P}|\eta > +\sin\theta_{P}|\eta^{\prime} > , \nonumber\\
&&  |\eta_{0} > = -\sin\theta_{P}|\eta > +\cos\theta_{P}|\eta^{\prime} > .  
\label{eta0}
\eea
Our basic idea is to compute the 
the $V\to \eta_{8}\gamma$  amplitude and to derive 
a sum rules relating the $\theta_{P}$ mixing angle to the 
measured $V\to\eta\gamma$ and $V\to\eta^{\prime}\gamma$ decay amplitude
by expressing the pure octet $\eta_{8}$ amplitude in terms
of the measured $\eta$ and $\eta^{\prime}$ amplitudes using
Eq. (\ref{eta0}). This is possible as the radiative decay branching 
ratios are currently known with  good accuracy \cite{PDG}. Defining 
the radiative decay 
electromagnetic form factor $V\to P$ by:
\be
<P(p_{P})|J^{\rm em}_{\mu}|V(p_{V})> = \epsilon_{\mu
  p_{P}p_{V}\epsilon_{V}}g_{VP\gamma}
\label{fVP}
\ee
where $J^{\rm em}_{\mu}$ the usual electromagnetic current in 
terms of quark field operators in $SU(3)$ space and $g_{VP\gamma} $ is the
on-shell $VP\gamma$ coupling constant with dimension  the inverse of energy.
The radiative decay
rates are then given by \cite{Ball}
\bea
&&\Gamma(V\to P\gamma)= \frac{\alpha}{24}g^{2}_{VP\gamma}
\Biggl(\frac{m_{V}^{2}-m_{P}^{2}}{m_{V}}\biggr)^{3}\nonumber\\
&&\Gamma(P\to V\gamma)= \frac{\alpha}{8}g^{2}_{VP\gamma}
\Biggl(\frac{m_{P}^{2}-m_{V}^{2}}{m_{P}}\biggr)^{3}
\label{VPrate}
\eea
For convenience, we give in Table. \ref{tab-1} the measured radiative
branching ratios together with the extracted coupling
constant $g_{VP\gamma}$ in  unit of $\rm GeV^{-1}$ and its theoretical value 
derived  either from an $SU(3)$ effective Lagrangian with nonet
symmetry for the $V\to\eta_{0}\gamma$ amplitude or from the 
quark counting rule with the coupling constant
$g_{VP\gamma}$ given in terms of the quark coupling constant  $g_{q}$,
($q=u,d,s$) for the magnetic  transition 
$(q\bar{q})(1^{-})\to (q\bar{q})(0^{-})\gamma$ \cite{Ball,Bramon,Escribano}.
The theoretical values for decay modes with $\eta$ in the final state is
obtained for the pure octet $\eta_{8}$( $\theta_{P}=0$) and  $SU(3)$
breaking effects  are taken into account with
$g_{s}=k\,g_{u}$ ($g_{d}=g_{u}$) for the magnetic transition 
$(q\bar{q})(1^{-})\to (q\bar{q})(0^{-})\gamma$ extracted  from the ratio of
the two measured $K^{*0}\to K^{0}\gamma$ to $K^{*\pm}\to K^{\pm}\gamma$
branching ratio with the magnetic coupling defined as \cite{Escribano}
\be
g_{K^{*0}K^{0}\gamma} = -g_{u}\frac{(1 + k)}{3}, \qquad g_{K^{*+}K^{+}\gamma} = g_{u}\frac{(2 - k)}{3}
\ee
where $k=\bar{m}/m_{s}$ is the constituent quark mass ratio \cite{Escribano}
in the quark model, but taken here as a parameter \cite{Escribano} 
and has a value $k=0.80\pm 0.06$ obtained from  the measured ratio \cite{PDG}
$BR(K^{*0}K^{0}\gamma)/BR(K^{*+}K^{+}\gamma)$ which is sensitive to $k$
. In addition to  $SU(3)$  and nonet symmetry breaking effects
in the magnetic coupling, as mentioned earlier, the 
renormalization of $K$ meson, $\eta_{8}$ and $\eta_{0}$ 
field operator in the $K^{*}\to K\gamma$,  $V\to\eta_{8}\gamma$ and 
in $V\to\eta_{0}\gamma$ amplitude  is given 
by the factor $f_{\pi}/f_{K}$, $f_{\pi}/f_{\eta_{8}}$, and 
$f_{\pi}/f_{\eta_{0}}$. In particular,  for $K^{*}\to K\gamma$ decay,
the factor $f_{\pi}/f_{K}$ with
$f_{K}= \rm 158\,\rm MeV$ and  the $SU(3)$ breaking 
factor $k$ are  needed to obtain agreement with experiments for the
computed $g_{K^{*}K\pi}$ coupling, as shown in  Table. \ref{tab-1} . 
Thus the corresponding $SU(3)$ and nonet symmetry breaking effect 
should  also be present in $V\to\eta_{8}\gamma$ and 
$V\to\eta_{0}\gamma$ amplitude. Here we take a 
recent chiral perturbation 
value $f_{\eta_{8}}=1.28\,f_{\pi}$,  $f_{\eta_{0}}=1.25\,f_{\pi}$
\cite{Kaiser,Leutwyler,Feldmann} (the value for $f_{\eta_{8}} $ is slightly
bigger  than the old value $ 1.25\,f_{\pi}$ in
\cite{Donoghue,Pham2,Gasser,Bijnens}) 
which produces a suppression factor $f_{\pi}/f_{\eta_{8}}=0.78$
and $f_{\pi}/f_{\eta_{0}}=0.80$
for  $V\to \eta_{8}\gamma$ and $V\to \eta_{0}\gamma$  relative to the
$V\to \pi^{0}\gamma$ amplitude, respectively. For the vector meson 
part of the amplitude, since  the isoscalar  vector mesons 
 exhibits an almost  ideal
mixing scheme, we use the quark flavor basis
to express the  $\omega$ and $\phi$ meson as linear 
superpositions of the
non strange $\omega_{0}=(u\bar{u} + d\bar{d})/\sqrt(2)$ and strange 
$\phi_{8}=s\bar{s}$ states with a mixing angle 
$\varphi_{V}=(3.2\pm 0.1)^{\circ}$
obtained from the $\omega\to \pi^{0}\gamma$ and $\phi\to \pi^{0}\gamma$
branching ratios \cite{Escribano}
\bea
&& |\omega> = \cos\varphi_{V}|\omega_{0} > - \sin\varphi_{V}|\phi_{8} > , \nonumber\\
&&  |\phi > = \sin\varphi_{V}|\omega_{0} > +\cos\varphi_{V}|\phi_{8} > .   
\label{v08}
\eea

\begin{table}[ht]
\begin{tabular}{|c|c|c|c|}
\hline
 Decay &$g_{VP\gamma}$, $\theta_{P}=0$, k=0.85&$g_{VP\gamma}(\rm
 exp.)$&\rm BR(exp) \cite{PDG} \\ 
\hline
\hline
$\rho^{\pm}\to \pi^{\pm}\gamma$&$(1/3)\,g_{u}$ &$0.72\pm 0.04$ &$ (4.5\pm 0.5)\times 10^{-4}$\\
$\rho^{0}\to \pi^{0}\gamma$&$(1/3)\,g_{u}$  &$0.83\pm 0.05$ & $ (6.0\pm 0.8)\times 10^{-4}$\\
$\rho^{0}\to \eta\gamma$&$0.58\,g_{u}\,(f_{\pi}/f_{\eta_{0}})$ &$1.59\pm 0.06$&$(3.00\pm 0.21)\times 10^{-4}$ \\
$ \omega\to \pi^{0}\gamma$&$0.99\,g_{u}$&$2.38\pm 0.03$ &$(8.92\pm 0.24)\%$ \\
$  \omega\to \eta\gamma$&$0.17\,g_{u}\, (f_{\pi}/f_{\eta_{0}})$&$0.45 \pm 0.02$  &$(4.6\pm 0.4)\times 10^{-4} $\\
 $  \phi\to \pi^{0}\gamma$&$0.06\,g_{u}$ &$0.13\pm 0.003$&$(1.26\pm 0.06)\times 10^{-3}$\\
$ \phi\to \eta\gamma$ &$0.47\,g_{u}\, (f_{\pi}/f_{\eta_{0}})$ &$0.71\pm 0.01 $&$(1.304\pm 0.025)\%$\\
$ \phi\to \eta^{\prime}\gamma$ &$-0.31\,g_{u}\, (f_{\pi}/f_{\eta_{0}}) $ &$-(0.72\pm 0.01) $&$(6.23\pm 0.21)\times 10^{-5}$\\
$ \eta^{\prime}\to \rho^{0}\gamma$ &$0.82\,g_{u}\, (f_{\pi}/f_{\eta_{0}}) $ &$1.35\pm 0.02 $&$(29.4\pm 0.9)\%$\\
$ \eta^{\prime}\to \omega\gamma$ &$0.29\,g_{u}\, (f_{\pi}/f_{\eta_{0}}) $ &$0.46\pm 0.02 $&$(3.02\pm 0.31)\%$\\
$K^{*\pm}\to K^{\pm}\gamma$&$0.38\,g_{u}\, (f_{\pi}/f_{K})$ &$0.84\pm 0.04$ &$ (9.9\pm 0.9)\times 10^{-4}$\\
$K^{*0}\to K^{0}\gamma$&$-0.62\,g_{u}\, (f_{\pi}/f_{K})$  &$-(1.27\pm 0.05)$ & $ (2.31\pm 0.20)\times 10^{-3}$\\
\hline
\end{tabular}
\caption{ Theoretical values for $V\to P\gamma$ with $\theta_{P}=0$,
k=0.85 together with the measured branching ratios and the extracted
$g_{VP\gamma}$}\label{tab-1}
\end{table}

As with the two-photon decay of $\eta$ meson \cite{Pham}, one can express
the pure octet $V\to \eta_{8}\gamma$ decay amplitude or the coupling constant 
$g_{V\eta_{8}\gamma}$ in terms of the physical $V\to \eta\gamma$
and $V\to \eta^{\prime}\gamma$ ($g_{V\eta\gamma}$ and 
$g_{V\eta^{\prime}\gamma}$) using Eq. (\ref{eta0}). Thus,
\be
S(V\to\eta\gamma)=g_{V\eta\gamma}\cos \theta_{P} + g_{V\to\eta^{\prime}\gamma}\sin \theta_{P}  = g_{V\eta_{8}\gamma}
\label{sr}
\ee
Since the  coupling constant $g_{V\eta_{8}\gamma}$ can be expressed in terms
of the theoretical value for $g_{V\pi^{0}\gamma}$ and SU(3) breaking 
parameters, Eq. (\ref{sr}) can be put into a more convenient form:
\be
S(V\to\eta\gamma)=(g_{V\eta\gamma}\cos \theta_{P} + g_{V\eta^{\prime}\gamma}\sin
\theta_{P})  = \biggl(\frac{g_{V\eta_{8}\gamma}}{g_{V\pi^{0}\gamma}}\biggr)_{\rm th.}g_{V\pi^{0}\gamma}
\label{sr1}
\ee
which becomes a sum rule relating the mixing angle
$\theta_{P}$ and the measured 
branching ratios of radiative decays involving $\eta$, $\eta^{\prime}$ and
$\pi^{0}$. The ratio 
$(\frac{g_{V\eta_{8}\gamma}}{g_{V\pi^{0}\gamma}})_{\rm th.}$
expresses the relative $V\to\eta_{8}\gamma$ with $SU(3)$ symmetry breaking
terms obtained experimentally from the measured $V\to\pi\gamma$
and $K^{*}\to K\gamma$ as explained above. The above sum rules 
allows a determination of the  
mixing angle $\theta_{P}$ with a minimum theoretical input like
$SU(3)$ breaking parameters which are  known to a good 
approximation. Using the computed values for $g_{V\eta_{8}\gamma}$
and the experimental values for $g_{V\eta\gamma} $ and 
$g_{V\eta^{\prime}\gamma} $ with $k=0.85$, $f_{\pi}/f_{K}=0.85$ and
 $f_{\pi}/f_{\eta_{8}}=0.78$  presented in Table.\ref{tab-1} , we find, for 
the l.h.s and r.h.s of Eq. (\ref{sr1}) for $\rho\to\eta\gamma$, $\omega\to\eta\gamma$
and $\phi\to\eta\gamma$ decays :
\bea
&&S(\rho\to\eta\gamma)=1.59\,\cos\theta_{P} + 1.35\,\sin\theta_{P}=1.12
\label{rho}\\
&&S(\omega\to\eta\gamma)=0.45\,\cos\theta_{P} + 0.46\,\sin\theta_{P}=0.31
\label{omega}\\
&&S(\phi\to\eta\gamma)=0.71\,\cos\theta_{P} - 0.72\,\sin\theta_{P}=0.88
\label{phi}
\eea

The above sum rules are very similar  to the sum rule we
obtained from the $\eta$ meson two-photon  decay:
\be
S(\eta\to\gamma\gamma)= 0.025\,\cos\theta_{P} + 0.03\,\sin\theta_{P}
=0.56\times 0.025(A_{\pi^{0}})
\label{eta2p}
\ee
where the measured two-photon decay amplitudes for $\eta$, $\eta^{\prime}$
and $\pi^{0}$ are numerically shown in  Eq. (\ref{eta2p}) above. From this
we obtain  a mixing angle of $\theta_{P}=-18.04^{\circ}$ mentioned
earlier.

It is clear from  Eqs. (\ref{rho}-\ref{phi}), that
$SU(3)$ breaking in the magnetic transition coupling  $g_{s}\not =
g_{u}$ and in
the $\eta_{8}$ decay constant $f_{\eta_{8}}\not = f_{\pi}$
in  the pure octet $\eta_{8}$ amplitude are not sufficient
to account for the measured $V\to \eta\gamma$ branching ratios which now need
a  negative value for  the mixing angle. If one neglects higher order
$SU(3)$ breaking effects and putting  $\cos\theta_{P}\approx 1$, one finds 
$\sin\theta_{P}\approx -0.34,-0.30,-0.23$, respectively, 
showing a  large first order $SU(3)$ symmetry breaking  in radiative decays. 
The exact solution of each of the above Eqs. (\ref{rho}-\ref{phi}), 
gives a  mixing angle 
$\theta_{P}=-(17.05\pm 4.4)^{\circ},-(15.51\pm 2.9)^{\circ},-(15.37\pm 2.1)^{\circ}$,  respectively
for $\rho\to\eta\gamma$, $\omega\to\eta\gamma$ 
and $\phi\to\eta\gamma$ radiative decays. These errors seem  a bit
large, especially for the value obtained from $\rho\to\eta\gamma$ decay, 
but are unavoidable, as  we are looking
for an $SU(3)$ breaking term affected by large experimental
error in the  difference of two measured quantities, the measured 
$V\to\eta\gamma$ on the l.h.s and the pure $\eta_{8}$ amplitude
given by the $V\to\pi^{0}\gamma$ amplitude on the r.h.s. of
Eqs. (\ref{rho}-\ref{phi}) . Thus to within
experimental error, it seems that our result could accommodate
the value  obtained from  nonet symmetry \cite{Kawarabayashi}
and from our previous value from $\eta$ meson two-photon decay
\cite{Pham}. We note that the determination of $\theta_{P}$
from $\rho\to\eta\gamma$ decay is less precise than the determination
by $\omega\to\eta\gamma$ , as the branching
ratios for $\rho\to\eta\gamma$ and $\rho\to\pi^{0}\gamma$ are known with
larger errors. Since the $\omega\to\pi^{0}\gamma$ branching ratio 
is currently known with an accuracy of about $3\%$, the main 
uncertainty in the determination of $\theta_{P}$
comes from $\omega\to\eta\gamma$ branching ratio  which is 
currently known with an accuracy at  $10\%$ level. Also some discrepancy with
the current data could  show up in new measurements of  light vector
meson radiative decays. In fact, the 
new KLOE \cite{KLOE} data, with the central
value of $\rm BR(\omega\to\pi^{0}\gamma)=8.09\%$ smaller by  $10\%$ than the
current PDG value \cite{PDG},  would imply   a mixing angle 
$\theta_{P}= -17.00^{\circ}$, slightly larger than
the  solution obtained here with the PDG value.

 In $\rho\to\eta\gamma$ and $\omega\to\eta\gamma$ decays, $SU(3)$ breaking  
 is due mainly to the factor $f_{\pi}/f_{\eta_{8}}$, thus mixing angle
obtained from   $\rho\to\eta\gamma$ and $\omega\to\eta\gamma$ decay
suffers from less theoretical uncertainties
than that from  $\phi\to\eta\gamma$ decay which is rather sensitive to
 the $SU(3)$ breaking effect for the $s$ quark magnetic coupling given by
 $g_{s}=k\,g_{u}$. To obtain  the value $-(15.31\pm 2.1)^{\circ}$ 
for $\phi\to\eta\gamma$ decay close to that from $\rho\to\eta\gamma$
and $\omega\to\eta\gamma$, we take $k=0.85$, a bit larger
than the value $k=0.80\pm 0.06$  from the
$K^{*}\to K\gamma$ branching ratios. This might not be a problem, since
there could be other $SU(3)$ breaking effects in $\phi\to\eta\gamma$
 not accounted for by $k$ alone and the $K^{*}\to K\gamma$   could have 
large experimental error as pointed out in \cite{Escribano} . Since 
\bea
&&\cos\varphi_{V}\,g_{\omega\eta_{8}\gamma} +
\sin\varphi_{V}\,g_{\phi\eta_{8}\gamma}= (\sqrt{3}/9)g_{u}\nonumber\\
&&\cos\varphi_{V}\,g_{\phi\eta_{8}\gamma} -
\sin\varphi_{V}\,g_{\omega\eta_{8}\gamma}= (2\sqrt{6}/9)g_{s}
\label{gv}
\eea
one could then try to 
eliminate this uncertainty by using, instead of the $\omega\to\eta\gamma$
amplitude alone, a linear combination  for an ideal mixing state, the 
$\omega_{0}\to\eta\gamma$ amplitude. We have
\be
S(\omega_{0}\to\eta\gamma)= \cos\varphi_{V}\,S(\omega\to\eta\gamma)+
\sin\varphi_{V}\,S(\phi\to\eta\gamma)\label{sot}
\ee
for  the ideal mixing $\omega_{0}$ state. We find
\be
S(\omega_{0}\to\eta\gamma)= 0.49\,\cos\theta_{P} + 0.42\,\sin\theta_{P}=0.36\label{sott}
\ee
which give $\theta_{P}=-(15.52\pm 3.3)^{\circ}$ consistent with the solution
obtained  with the $\omega\to\eta\gamma$ amplitude alone. Similarly, the
linear combination amplitude for the pure $s\bar{s}$ state 
depends only on the $s$ quark magnetic coupling and  is given by
\be
S(\phi_{8}\to\eta\gamma)= 0.68\,\cos\theta_{P} - 0.75\,\sin\theta_{P}=0.85
\ee
which gives a solution $\theta_{P}=-(15.37\pm 3.9)^{\circ}$, also consistent
with the all the solutions obtained above.

 We have obtained the mixing angle $\theta_{P}$ 
 by using the sum rules for $V\to\eta\gamma$ alone. Since the derivative 
mixing term  affects essentially the $V\to\eta^{\prime}\gamma$, by using
both sum rules for $V\to \eta\gamma$ and  $V\to\eta^{\prime}\gamma$
and nonet symmetry for the pure $SU(3)$ singlet $V\to\eta_{0}\gamma$,
one would be able to determine both $\theta_{P}$ and $d$. The two sum rules, 
similar to Eq. (\ref{sr1}), are then
\bea
&&S(V\to\eta\gamma)=g_{V\eta\gamma}\cos \theta_{P} + 
(g_{V\eta^{\prime}\gamma}-d\,g_{V\eta\gamma}) \sin
\theta_{P}  = \biggl(\frac{g_{V\eta_{8}\gamma}}{g_{V\pi^{0}\gamma}}\biggr)_{\rm th.}g_{V\pi^{0}\gamma}\label{sr11}\\
&&S(V\to\eta^{\prime}\gamma)=(g_{V\eta^{\prime}\gamma}-d\,g_{V\eta\gamma})\cos \theta_{P} - 
g_{V\eta\gamma}\sin
\theta_{P} = \biggl(\frac{g_{V\eta_{0}\gamma}}{g_{V\pi^{0}\gamma}}\biggr)_{\rm th.}g_{V\pi^{0}\gamma}\label{sr12}
\eea
neglecting second order term $d\,g_{V\eta^{\prime}\gamma} $
in Eq. (\ref{sr11}). With $k=0.85$, $f_{\pi}/f_{K}=0.85$, 
$f_{\pi}/f_{\eta_{8}}=0.78$, $f_{\pi}/f_{\eta_{0}}=0.80$ 
and the nonet symmetry value for
$V\to\eta^{0}\gamma$ shown in Table.\ref{tab-1}, we have, for 
the l.h.s and r.h.s of Eq. (\ref{sr1}) for $\rho\to\eta,\eta^{\prime}\gamma$
in Eq. (\ref{sr11}) and Eq. (\ref{sr12})
\bea
&&S(\rho\to\eta\gamma)=1.59\,\cos\theta_{P} + (1.35-1.59\,d)\sin\theta_{P}=1.12
\nonumber\\
&&S(\eta^{\prime}\to\rho\gamma)=(1.35 -1.59\,d)\cos\theta_{P} -1.59\,\sin\theta_{P}=1.63
\label{rho2}
\eea
Similarly, for $\omega$ and $\phi$, we have:
\bea
&&S(\omega\to\eta\gamma)=0.45\,\cos\theta_{P} + (0.46-0.45\,d)\sin\theta_{P}=0.30
\nonumber\\
&&S(\eta^{\prime}\to\omega\gamma)=(0.45 -0.46\,d)\cos\theta_{P} -0.45\,\sin\theta_{P}=0.55
\label{omega2}
\eea
and 
\bea
&&S(\phi\to\eta\gamma)=0.71\,\cos\theta_{P} + (-0.72-0.71\,d)\sin\theta_{P}=0.88
\nonumber\\
&&S(\phi\to \eta^{\prime}\gamma)=(-0.72 -0.71\,d)\cos\theta_{P} -0.71\,\sin\theta_{P}=-0.59
\label{phi2}
\eea

The solutions of the above coupled equations then give 
$\theta_{P}$,  $-(18.76\pm 4.4)^{\circ},-(15.81\pm
3.1)^{\circ},-(13.83\pm 2.1)^{\circ} $
and  $d$, $0.10\pm 0.05$, $0.02\pm 0.03$ , $0.08 \pm 0.03$ for $\rho$,  $\omega$ and $\phi$
respectively.  
For the ideal $\omega_{0}$ and  $\phi_{8}$ state , the sum rules $S(\omega_{0}\to\eta,\eta^{\prime}\gamma)$,
 $S(\phi_{8}\to\eta,\eta^{\prime}\gamma)$ give 
$\theta_{P}$, $-(16.03\pm 3.5)^{\circ}$,  $-(15.37\pm 3.8)^{\circ}$ and 
$d$, $0.03\pm 0.03$,  $0.09\pm 0.03 $  close to the values for 
 $\omega$ and $\phi$ state.
Our value for $d$ is somewhat smaller than the corresponding 
 value of $0.14-0.16$ for $\sin(\theta_{0}-\theta_{8})$ obtained
 to first order in  $SU(3)$  breaking in the two-angle mixing scheme for the
 pseudo-scalar meson decay constants \cite{Kaiser}. The value of $d$ for 
$\eta^{\prime}\to \omega\gamma$ is rather small, but with large
experimental errors. To reduce these errors, one could
determine $d$ directly from the 
$V\to\eta^{\prime}\gamma$ amplitudes with the nonet symmetry
$V\to\eta^{0}\gamma$ amplitude 
 and a mixing angle of $-18^{\circ}$ obtained from 
nonet symmetry for the momentum-independent mixing mass term 
 \cite{Kawarabayashi}. We find  $d$ :  $0.09\pm
 0.04$, $0.06\pm 0.02$ and $0.15 \pm 0.03$ for $\rho$,  $\omega$ and $\phi$
respectively, comparable to the chiral perturbation
 results \cite{Kaiser}. We note that  a mixing angle of $-22^{\circ}$
could produce  a larger $d$ :  $0.16\pm 0.04$, $0.13\pm 0.02$ and  $0.21 \pm
0.03$ for $\rho$, $\omega$ and $\phi$ , respectively, corresponding 
to a small $\theta_{0}$ found in \cite{Escribano1}.

The above values for $\theta_{P}$ are quite close to the values
obtained  from the sum rules  $S(V\to\eta\gamma$) alone. Then,
considering current theoretical and experimental
uncertainties, one could just use the sum rules with $\eta$ alone to obtain
the new mixing angle for processes with $\eta$ meson without involving 
the pure singlet $V\to\eta_{0}\gamma$, but for processes with $\eta^{\prime}$
one need to know $d$ either from the two sum rules with theoretical
input for the pure singlet $\eta_{0}$ amplitude or from some other
method. 

\bigskip

   In conclusion, we have derived sum rules relating the new 
$\eta-\eta^{\prime}$ mixing angle to the measured $V\to\eta\gamma$ 
and $V\to\eta^{\prime}\gamma$
decay amplitude which allows a determination of the mixing angle using only
the measured radiative decay branching ratios. With only the $\eta$ sum rules,
we  find $\theta_{P}$ in the range  $-14^{\circ}$  to $-17^{\circ}$
 within an error of $(2.1-4.4)^{\circ}$, with two  $\eta,\eta^{\prime}$ 
sum rules,  we find a similar value   in the 
range $-14^{\circ}$  to $-19^{\circ}$ and an evidence for the 
momentum-dependent mixing mass term $d$ in $\rho$ and $\phi$ 
radiative decays. We also obtain large $d$ in $\eta^{\prime}\to
\omega\gamma$ using  nonet symmetry,  comparable to $d$ from $\rho$ and $\phi$
decay. More precise vector meson radiative decay measured branching ratios
and higher order $SU(3)$ breaking effects could bring these 
extracted values for  $\theta_{P}$ closer and give us  a  better 
determination of the momentum-dependent mixing term $d$ which 
is needed in  processes with $\eta^{\prime}$.

\begin{center}
{\bf Acknowledgments} \end{center}

This work was supported in part by the EU 
contract No. MRTN-CT-2006-035482, "FLAVIAnet".


\end{document}